\documentclass[12pt]{article}
\usepackage{latexsym}
\newcommand{\beq}{\begin{equation}}
\newcommand{\eeq}[1]{\label{#1}\end{equation}}
\newcommand{\bea}{\begin{eqnarray}}
\newcommand{\eea}[1]{\label{#1}\end{eqnarray}}

\begin{document}
\hfill 
\begin{flushright}
NYU-TH/00/11/03 \\
YITP-00-78 \\
November, 2000\\
hep-th/0011278
\end{flushright}

\vspace{20pt}

\begin{center}
{\textbf{NO VAN DAM-VELTMAN-ZAKHAROV DISCONTINUITY \\
FOR SUPERGRAVITY IN ADS SPACE}}
\end{center}

\vspace{6pt}

\begin{center}
{\bf P.A. Grassi$^{a}$} and  {\bf P. van Nieuwenhuizen$^{b}$} \vspace{20pt}

\textit{(a) Department of Physics, NYU, 4 Washington Pl, New York NY 10003, USA. \\ 
(b) C.N. Yang Institute for Theoretical Physics, Stony Brook, NY 11794-3840, USA.}

\end{center}

\vspace{12pt}

\begin{center}
\textbf{Abstract }\end{center}

Adding explicit mass terms for the spin 2 and spin 3/2 field of $N=1$ anti-de Sitter supergravity, the limit 
$M^2 \rightarrow 0$  for these mass terms is smooth: there is no van Dam-Veltman-Zakharov mass discontinuity 
in the propagators when the cosmological constant is non-vanishing.

\vspace{4pt} {\small \noindent

\vfill\eject 
\noindent

Recently Kogan {\it et al.}~\cite{kogan} and Porrati~\cite{porr} have shown 
that the mass discontinuity arising in the massless limit of massive gravity theories 
is peculiar to Minkowski space and does not arise in anti-de Sitter spaces. 
A similar result for de Sitter spaces was obtained by Higuchi~\cite{higuchi}, but in this case 
there are fewer physical applications (there are no unitary representations~\cite{porr}, and supergravity does
not exist in de Sitter space~\cite{su_de}).  

The action they consider is a sum of the Einstein action, a cosmological
term $\Lambda$ and the spin 2 Fierz-Pauli~\cite{pf} mass term 
\begin{eqnarray}
  \label{eq_0}
  {\cal L}_{E} &=& 
\frac{\sqrt{-{\rm det}(g+h)}}{2}\left[ R(g+h) - 2 \,\Lambda  + h_{\mu\nu} T^{\mu\nu}\right] 
 +  {\cal L}^{(2)}_{(mass)}\,, \nonumber\\
  \label{eq_1}
   {\cal L}^{(2)}_{(mass)} &=& - \frac{\sqrt{-g}\, M^2}{8}  
\Big( h_{\mu\nu} h_{\rho\sigma} g^{\mu\rho} g^{\nu\sigma} - 
 h_{\mu\nu} h_{\rho\sigma} g^{\mu\nu} g^{\rho\sigma} \Big)\,. 
\end{eqnarray}
Here $g_{\mu\nu}$ is the background metric for the Einstein space.
All indices are raised, lowered and contracted with the 
background metric $g_{\mu\nu}$, and the 
quantum gravitational field $h_{\mu\nu}$ is coupled to an external 
source $T_{\mu\nu}$ which is covariantly conserved in the background metric.

The propagator with both $\Lambda$ and $M$ non-vanishing has only a pole 
at $\nabla^2 = M^2 - 2 \Lambda$ whose residue is~\cite{kogan,porr}
\begin{equation}
  \label{eq_2}
 T^{\mu\nu} T_{\mu\nu} + T \left[ \frac{\Lambda - M^2}{ 3 M^2 - 2 \Lambda} \right] T\,,
\end{equation}
where $T$ denotes the trace of $T^{\mu\nu}$. For $\Lambda \rightarrow 0$ at fixed $M^2$, one finds 
$T^{\mu\nu}T_{\mu\nu} - \frac{1}{3} T^2$ which is the residue of a massive spin 2 particle. 
This residue is positive and thus tree level unitarity is preserved~\cite{pvn}. 
However, for $M^2 \rightarrow 0$  at fixed $\Lambda$ one finds instead 
$T^{\mu\nu}T_{\mu\nu} - \frac{1}{2} T^2$ which is the residue of a massless 
spin 2 particle, and which also satisfies the tree level unitarity conditions~\cite{pvn}. 
The fact that $T^{\mu\nu}T_{\mu\nu} - \frac{1}{3} T^2$ is discreetly different 
from $T^{\mu\nu}T_{\mu\nu} - \frac{1}{2} T^2$ is the well-known van Dam-Veltman-Zakharov 
(vDVZ) mass-discontinuity~\cite{vvdz}. As observed in \cite{kogan,porr}, the discontinuity 
is an accident in Minkowski space, but the limit $M\rightarrow 0$ is smooth in anti-de
Sitter space  ($\Lambda <0$)~\cite{kogan,porr} and de Sitter space ($\Lambda >0$)~\cite{higuchi}.

In this note we extend this result to spin $3/2$ Rarita-Schwinger
fields. Then we put the results into the context of supergravity
with super-cosmological constant. Finally, 
we also generate a mass term through a generalized St\"uckelberg 
formalism. 

The action for a massive spin 3/2 field in curved space is~\cite{rep} 
\begin{equation}
  \label{eq_3}
  {\cal L}^{3/2} = - \frac{e}{2} \bar\Psi_\mu \gamma^{\mu\rho\sigma} \nabla_\rho  \Psi_\sigma
                              + \frac{M}{2}   \bar\Psi_\mu \gamma^{\mu\nu} \Psi_\nu
                              + \bar\Psi^\mu J_\mu\,,
\end{equation}
where $\gamma^\mu = e^\mu_m \gamma^m$ with constant $\gamma^m$, $e = {\rm det} (e_\mu^m)$, 
$\gamma^{\mu\nu} = \frac{1}{2} \left[ \gamma^\mu, \gamma^\nu\right]$ and 
$\gamma^{\mu\nu\rho}$ is the totally antisymmetric part of 
$\gamma^{\mu}\gamma^{\nu}\gamma^{\rho}$. Further, $\Psi_\mu$ is a Majorana spinor , so 
$\bar\Psi_\mu = \Psi^T_\mu C$ with $C$ the charge conjugation matrix, and 
\begin{equation}
\label{eq_4}
\nabla_\rho \Psi_\sigma = \partial_\rho \Psi_\sigma + 
\frac{1}{4} \omega_{\rho}^{~mn}(e) \gamma_{\,mn} \Psi_{\rho}
-\Gamma_{\rho \sigma}^{~~\tau} \Psi_\tau\,,
\end{equation}
where $\Gamma_{\rho \sigma}^{~~\tau}$ is the Christoffel connection and 
$\omega_{\rho}^{~mn}(e)$ the spin connection. The source $J_\mu$ is an external source, which 
we take to be covariantly conserved; this is necessary in the massless theory ~\cite{rep}, 
but not necessary 
for the massive theory, although Kaluza-Klein theories automatically couple massive fields 
to conserved sources~\cite{kogan}. 

The field equations read 
\begin{equation}
  \label{eq_5}
   \gamma^{\mu\rho\sigma} \nabla_\rho  \Psi_\sigma - M \gamma^{\mu\nu} \Psi_\nu = J^\mu\,.
\end{equation}
Using $\gamma^{\mu\rho\sigma} = \gamma^{\mu}\gamma^{\rho}\gamma^{\sigma} - 
g^{\mu\rho} \gamma^\sigma  - g^{\rho\sigma} \gamma^\mu + g^{\mu\sigma} \gamma^\rho$, 
we obtain
\begin{equation}
  \label{eq_6}
   \gamma_\mu \not\! \nabla \gamma\cdot \Psi 
- \nabla_\mu \gamma\cdot \Psi 
- \gamma_\mu \nabla \cdot \Psi  
+ \not\! \nabla \Psi_\mu - M \gamma_{\mu\nu}  \Psi^\nu = J_\mu\,,
\end{equation}
where we used that the derivative $\nabla_\rho$ in~(\ref{eq_4}) commutes with $\gamma_\sigma$, 
so $[\nabla_\rho, \gamma_\sigma] =0$. To obtain explicit expressions for the lower spin parts 
$\gamma\cdot \Psi$ and $\nabla\cdot\Psi$ in terms of $J^\mu$, we contract the field equations 
with $\nabla^\mu$ and $\gamma^\mu$, respectively. This yields
\begin{eqnarray}
  \label{eq_7.1}
&&  \not\!\nabla   \not\!\nabla (\gamma\cdot\Psi) - \nabla^2 (\gamma\cdot \Psi) - 
\not\!\nabla (\nabla\cdot \Psi) + \nabla_\mu (\not\!\nabla\Psi^\mu) - 
M \gamma^{\mu\nu}  \nabla_\mu \Psi_\nu = \nabla \cdot J = 0 \,, 
\\ \label{eq_7.2}
&& 2\, \not\!\nabla (\gamma\cdot\Psi) - 2 \, \nabla\cdot\Psi - 3\, M  \gamma\cdot\Psi = 
\gamma \cdot J\,,
\end{eqnarray}
where we used $\gamma^\mu \not\!\nabla \Psi_\mu = - \not\!\nabla \gamma \cdot \Psi + 2 \nabla \cdot \Psi$ in~(\ref{eq_7.2}).
In~(\ref{eq_7.1}), we used 
\begin{equation}
  \label{eq_8}
  \nabla_\mu \not\!\nabla \Psi^\mu = \not\!\nabla \nabla \cdot \Psi + 
\gamma^\nu [ \nabla_\mu,\nabla_\nu] \Psi^\mu\,. 
\end{equation}

We now need some properties of gravitationally covariant derivative for spinors. First of all, 
from the vielbein ``postulate'' 
$\nabla_\nu e^m_\rho \equiv \partial_\nu e^m_\rho - \Gamma_{\nu\rho}^{~\sigma} e_{\sigma}^m 
+ \omega_{\nu~~~n}^{~m} e^n_\rho =0$ we obtain $[\nabla_\mu,\nabla_\nu] e^{~m}_\rho = 0$, which relates the curvature in
term  of Christoffel symbols to the curvature in term of the spin connection
\begin{equation}
  \label{eq_9}
  - R_{\mu\nu\rho}^{~~~~\sigma}(\Gamma) e_{\sigma}^m + R_{\mu\nu~ n}^{~~m}(\omega) e_{\rho}^n = 0\,, 
\end{equation}
where 
\begin{eqnarray}
  \label{eq_10}
  R_{\mu\nu\rho}^{~~~~\sigma}(\Gamma) 
&=& \partial_\mu \Gamma_{\nu\rho}^{~~\sigma} + \Gamma_{\mu\tau}^{~~\sigma} \Gamma_{\nu\rho}^{~~\tau} - 
(\mu \leftrightarrow \nu)\,, \nonumber \\
R_{\mu\nu~n}^{~~m}(\omega) 
&=& \partial_\mu \omega_{\nu~n}^{~m} + \omega_{\mu~k}^{~m} \, \omega_{\nu~n}^{~k} - 
(m \leftrightarrow n)\,.
\end{eqnarray}
We define anti-de Sitter space by 
\begin{eqnarray}
  \label{eq_11}
   R_{\mu\nu\rho\sigma}(\Gamma)  &=& \frac{\Lambda}{3} \left( g_{\mu\sigma} g_{\nu\rho} - g_{\mu\rho} g_{\nu \sigma} \right) \,,
\nonumber \\
   R_{\mu\sigma}(\Gamma) &=& R_{\mu\nu\rho\sigma}(\Gamma)  g^{\nu\rho} = \Lambda g_{\mu\sigma} \,.
\end{eqnarray}

Next we recall that, although the Christoffel symbols cancel in the curls 
$\nabla_\rho \Psi_\sigma - \nabla_\sigma \Psi_\rho$ in the action, they are still present 
in $\nabla_\mu \not\!\nabla \psi^\mu$. Hence, 
\begin{equation}
  \label{eq_12}
  [\nabla_\mu, \nabla_\nu] \Psi^\mu = -   R_{\mu\nu}^{~~~\mu\tau}(\Gamma) \Psi_\tau + 
 \frac{1}{4}  R_{\mu\nu}^{~~m n}(\omega) \gamma_{\, mn} \Psi^\mu = 
\Lambda \Psi_\nu + \frac{\Lambda}{6} \gamma_{\mu\nu}\Psi^\mu\,.
\end{equation}
Similarly, 
\begin{eqnarray}
  \label{eq_13}
&&\not\!\nabla \not\!\nabla (\gamma\cdot \Psi) = 
\nabla^2 (\gamma\cdot \Psi) + \frac{1}{2} \gamma^\mu \gamma^\nu [\nabla_\mu,\nabla_\nu] \gamma\cdot \Psi = \nonumber \\
&&\mbox{}~~~~\nabla^2 (\gamma\cdot \Psi) +  \frac{1}{8} \gamma^\mu \gamma^\nu  
R_{\mu\nu}^{~~m n}(\omega) \gamma_{\, mn}  (\gamma\cdot \Psi)  = 
\nabla^2 (\gamma\cdot \Psi)  - \Lambda  (\gamma\cdot \Psi) \,.
\end{eqnarray}

With these results, the contracted equations reduce to 
\begin{eqnarray}
  \label{eq_14.1}
  && \frac{\Lambda}{2} \, \gamma\cdot \Psi + M \Big( \not\!\nabla (\gamma\cdot \Psi) - \nabla \cdot \Psi \Big) = 0\,, \\
\label{eq_14.2}  && 2\,\not\!\nabla(\gamma\cdot \Psi) - 2 \, \nabla\cdot\Psi - 3\, M \,\gamma\cdot \Psi = \gamma \cdot J\,.
\end{eqnarray}

The first equation allows to express $\nabla \cdot \Psi$ in terms of 
$\gamma \cdot \Psi$, and substituting the result into~(\ref{eq_14.1}) yields an expression for
$\gamma\cdot \Psi$ in terms of $\gamma \cdot J$:
\begin{eqnarray}
&&\nabla \cdot \Psi = \left( \frac{\Lambda}{2 \, M} + \not\!\nabla \right) \gamma \cdot \Psi\,, \label{eq_15.1} \\
&& \gamma \cdot \Psi = -\left( \frac{\Lambda}{M} + 3\, M \right)^{-1} \gamma \cdot J \,.
\end{eqnarray}
Substituting these results into the field equation (\ref{eq_6}) leads to the propagator 
\begin{equation}\label{eq_16}
\bar J^\mu \Psi_\mu = 
\bar J^\mu  \left\{\frac{\left( \Lambda + 3\, M^2\right)^{-1}}{\not\!\nabla + M} 
\left[ - M \, \nabla_\mu \gamma_\nu - \left(\frac{\Lambda}{2} + M^2 \right) \gamma_\mu \gamma_\nu 
+ g_{\mu\nu} (\Lambda + 3\, M^2) 
  \right] \right\} J^\nu\,.
\end{equation}\label{eq_17}

For fixed $\Lambda$ but $M \rightarrow 0$ we obtain 
\begin{equation}
\bar J^\mu \frac{1}{\not\!\nabla} \left[ g_{\mu\nu} - \frac{1}{2} \gamma_\mu \gamma_\nu \right] J^\nu\,.
\end{equation}
This is  the propagator for a massless spin 3/2 particle. In flat space 
the residue can be written as  
$\bar J^\mu \left( \eta_{\mu\nu} \not\!\partial + \frac{1}{2}\gamma_\mu \not\!\partial\,
\gamma_\nu \right) J^\nu$ which is positive definite, 
and the theory is thus unitary at tree level~\cite{rep}. On the other hand , for fixed $M$ but $\Lambda \rightarrow 0$, 
we obtain 
\begin{equation}\label{eq_19}
\bar J^\mu \frac{1}{\not\!\nabla + M}
\left[ - \frac{1}{3\, M} \nabla_\mu \gamma_\nu + g_{\mu\nu} - \frac{1}{3} \gamma_\mu \gamma_\nu \right] J^\nu\,.
\end{equation}
In the flat space $\nabla_\mu$ commutes with $~(\not\!\!\nabla + M)^{-1}$ and annihilates on $\bar J^\mu$, and
we  find the propagator of a massive spin 3/2 particle in flat space~\cite{rep}
 \begin{equation}\label{eq_20}
\bar J^\mu \frac{1}{\Box - M^2}
\left[ \eta_{\mu\nu} \left( \not\!\partial -M\right) + \frac{1}{3} \gamma_\mu 
\left( \not\!\partial + M\right) \gamma_\nu \right] J^\nu\,.
\end{equation}
Also this propagator has a positive definite residue, and thus satisfies tree level unitarity~\cite{rep}. 

The fact that $\eta_{\mu\nu}\not\!\partial + \frac{1}{2} \gamma_\mu \not\!\partial\,\gamma_\nu$ is discreetly 
different from $ \eta_{\mu\nu}\left( \not\!\partial -M\right) + \frac{1}{3} \gamma_\mu 
\left( \not\!\partial + M\right) \gamma_\nu$ for $M\rightarrow 0$ is a van Dam-Veltman-Zakharov mass discontinuity, but in
curved space  with non-vanishing $\Lambda$ the limit $M\rightarrow 0$ is smooth. Thus spin 3/2 fields 
and spin 2 fields behave
the same way as far as the limit $M\rightarrow 0$ is concerned, which are reasons to turn to supergravity. 

First we consider what happens if one does not take the flat-space limit. 
In curved space, $\nabla^\mu$ does, of course, not commute with 
$(\not\hspace{-.15cm}\nabla + M)^{-1}$. It is possible to rewrite 
$(\not\hspace{-.15cm}\nabla + M)^{-1}$ as 
$\left[ (\not\!\nabla - M) (\not\!\nabla + M)\right]^{-1}(\not\hspace{-.15cm}\nabla - M)
$. Furthermore, the operator $\left[ (\not\!\nabla - M) (\not\!\nabla + M)\right]^{-1} $ is the 
inverse of 
$\left[ (\not\!\nabla - M) (\not\!\nabla + M)\right]$ and the latter satisfies 
\begin{equation}\label{eq_21}
\left[ (\not\!\nabla - M) (\not\!\nabla + M), \nabla_\mu\right] \gamma\cdot J= \Lambda \nabla_\mu (\gamma\cdot J)\,.
\end{equation}
Acting on a vector spinor $J_\mu$, it reads 
\begin{eqnarray}\label{eq_22}
\left[ (\not\!\nabla - M) (\not\!\nabla + M)\right]  J_\nu &=& \left( \nabla^2 - M^2 \right) J_\nu + 
\frac{1}{2} \gamma^\rho \gamma^\sigma \left[ - 
R_{\rho\sigma\nu}^{~~~~\tau}(\Gamma) J_\tau + \frac{1}{4} R_{\rho\sigma}^{~~~m n}(\omega) \gamma_{mn}  J_\nu 
\right]  \nonumber \\
&=&\left( \nabla^2 - M^2  - \Lambda \right) J_\nu + \frac{\Lambda}{3} \gamma_{\nu\mu} J^\mu \,.
\end{eqnarray}
The terms with $\Lambda$ are similar to the terms with the $\Lambda$ in the Lichnerowicz operator
\footnote{To check the constants, 
note that if $T_{\mu\nu} = g_{\mu\nu} A$, one has $- \Delta^{(2)}_L (g_{\mu\nu}A) =
\Box A$.} 
$\Delta^{(2)}_L$ acting on $T^{\mu\nu}$
\begin{equation}\label{eq_23}
- \Delta^{(2)}_L T^{\mu\nu}= \Box T^{\mu\nu} + \frac{2}{3} \Lambda g_{\mu\nu} T - \frac{8}{3} T^{\mu\nu}\,. 
\end{equation}
 One can decompose $ \Delta^{(2)}_L$ into a traceless part and a trace, and invert each part separately, 
just as one decomposes the propagator for gauge fields into a transversal part and a longitudinal part. 
(One can then prove that the latter does not renormalize). 
Similarly, one could decompose the spin 3/2 propagator ~\cite{bbgr}.

We now put these results into the context of supersymmetry. For $N=1$ supergravity, one can 
add a cosmological constant and still preserve local supersymmetry~\cite{nsc}. The action and 
transformation rules read
\begin{eqnarray}\label{eq_24}
&&{\cal L} = {\cal L}_E + {\cal L}_{RS} + e \left( \frac{6\alpha}{\kappa^4} + \frac{\alpha}{\kappa} \bar\Psi_\mu \gamma^{\mu\nu}
\Psi^\nu\right) 
\nonumber \\
&&\delta \Psi_\mu = \frac{1}{\kappa} \nabla_\mu \epsilon +\frac{\alpha}{\kappa^2} \gamma_\mu \epsilon\,;~~~~~
\delta e_\mu^{~m} = \frac{\kappa}{2}\, \bar\epsilon \,\gamma^m \psi_\mu\,,
\end{eqnarray}
with $\alpha$ a free constant. We note that the apparent mass term 
$\frac{\alpha}{\kappa}\bar\Psi_\mu \gamma^{\mu\nu}\Psi^\nu$ 
has the same form as the explicit mass term in~(\ref{eq_3}), 
but for spin 2 fields the expansion of the cosmological term leads to
a different quadratic mass term than the Fierz-Pauli mass term
\begin{eqnarray}\label{eq_25}
&&\frac{1}{\kappa^4} \sqrt{-{\rm det}(g +h)} = \frac{\sqrt{-g}}{\kappa^4} \left[ 1 + \frac{1}{2} h - \frac{1}{4} \left(
h_{\mu\nu} h^{\mu\nu} - 
\frac{1}{2}h^2 \right) + \dots \right] \,,\nonumber \\
  && {\cal L}^{(2)}_{(mass)} = - \frac{\sqrt{-g}\, M^2}{8}  
\Big( h_{\mu\nu} h^{\mu\nu} - h^2  \Big)\,.
\end{eqnarray}
where all the contractions are performed with the background metric $g_{\mu\nu}$. 
In flat space the massive free spin 2 and spin 3/2 system with Fierz-Pauli mass terms
\begin{eqnarray}\label{eq_pf}
{\cal L}_{FP} &=& - \frac{1}{2} \left(\partial_\lambda h_{\mu\nu}\right)^2 + 
\left(\partial^\nu h_{\mu\nu} \right)^2 - \partial^\mu h \,\partial^\nu h_{\mu\nu} + \frac{1}{2} 
\left(\partial^\mu h\right)^2 - \frac{1}{2} \bar\Psi_\mu 
\gamma^{\mu\rho\sigma} \partial_\rho \Psi_\sigma \nonumber \\
&-& \frac{M^2}{2} \Big( h^2_{\mu\nu}  - h^2  \Big) + \frac{M}{2} \bar\Psi_\mu \gamma^{\mu\nu} \Psi_\nu 
\end{eqnarray}
is not invariant under the rigid susy transformation rules for 
anti-de Sitter supergravity
\begin{equation}\label{eq_27}
\delta h_{\mu\nu} = \frac{1}{2} \left( \bar\epsilon \gamma_\mu \Psi_\nu + \bar\epsilon \gamma_\nu \Psi_\mu \right), 
~~~~~\delta\Psi_\mu = \frac{1}{4} \hat\omega_\mu^{~~mn} \gamma_{m n} + \alpha M\, h_{\mu\nu} \gamma^\nu \epsilon 
+ \beta M \, h \gamma_\mu \epsilon
\end{equation}
where $\hat\omega_{\mu m n} =\frac{1}{8} \left(- \partial_m h_{\mu n} + \partial_n h_{\mu m}\right)$ 
is the linearized spin connection.  

For the massless case, we can promote the rigid susy to a local susy by using the Noether method and fusing the local 
symmetry $\delta \Psi_\mu = \partial_\mu \eta$  of the linearized spin 3/2 action with 
the linearized rigid susy transformation rules. The result is
$N=1$  supergravity~\cite{rep}. However, to repeat this procedure for  the massive theory one runs into a wall. 
This is not surprising: a massive spin 2 has 5 degrees of freedom and a massive spin 3/2 has 4 degrees of freedom. The
massive representation of $N=1$ susy with spin 2 contains one massive  graviton, two gravitinos (one complex gravitino) 
and one massive real vector field. The degrees of freedom now match: $5+3=4+4$, and one can begin with a linearized 
rigid susy system, and use the Noether method to construct the non-linear theory. 

One must then not only add a cubic term to the action and
quadratic terms to the transformation rules as usual in the Noether method, but also a term linear in
$h$ to the action, and a field-independent local term $\delta \Psi_\mu = \gamma_\mu \epsilon$ to the transformation rules. 
The result is $N=2$ supergravity with a super-cosmological term~\cite{sc}.

As a final remark, we discuss how to generate masses
for the spin 2 Fierz-Pauli action~(\ref{eq_pf}) 
and for the spin 3/2 Rarita-Schwinger action (\ref{eq_3}) by the St\"uckelberg formalism~\cite{stuck,g}. 
As is well-known, in the case of abelian gauge theories, 
the St\"uckelberg formalism provides a gauge invariant procedure to 
describe  a massive vector field. Due to the absence of couplings with the Faddeev-Popov 
ghosts the theory is renormalizable
and the  limit $M\rightarrow 0$ is smooth to all orders. The same technique can be used for the linearized 
Fierz-Pauli action~(\ref{eq_pf}) by introducing an auxiliary  spin 1 field 
and an auxiliary spin zero field~\cite{higuchi}. It turns out
that  in flat space the action of the gauge field and of the scalar field
are the Maxwell action and the Klein-Gordon action if one chooses the mass terms 
appropriately. This implies that 
the action with the auxiliary fields is ghost free. 
 
For the spin 3/2 case,
we introduce an auxiliary Majorana spin 1/2 field $\lambda$ which transforms in the following way
\begin{equation}\label{stu_1}
\delta \Psi_\mu = \partial_\mu \eta \,, ~~~~\delta \lambda = \eta\,.
\end{equation}
Therefore, the combination $\Psi_\mu - \partial_\mu \lambda $ is gauge invariant. 
The spin 3/2 mass term can be written in a gauge invariant way as
\begin{equation}\label{stu_2}
{\cal L}^{(3/2)}_{mass} =
\frac{M}{2}  \left(\bar\Psi_\mu - \partial_\mu \bar\lambda \right)\gamma^{\mu\nu} 
\left(\Psi_\nu - \partial_\mu \lambda \right) = 
\frac{M}{2} \left( \bar\Psi_\mu \gamma^{\mu\nu}  \Psi_\nu - 2 \bar\Psi_\mu \gamma^{\mu\nu} \partial_\mu \lambda \right) 
\,.
\end{equation}
The absence of ghosts in the action is due to the cancellation of higher derivatives 
for the spin 1/2 fields: due to the antisymmetry of the tensor structure $\gamma^{\mu\nu}$ 
the double derivative term cancels. 
From this point of view, it becomes clear why a mass term $\bar\Psi_\mu\Psi^\mu$ is not allowed: it would lead to a 
higher-derivative action $\partial_\mu \bar\lambda \partial^\mu \lambda$ for the spin 1/2 field.
A suitable choice of the gauge fixing
is needed  in order to remove the couplings $\bar\Psi_\mu \gamma^{\mu\nu} \partial_\mu \lambda$, and this generates the 
Dirac action for the spin 1/2 field. Further details will be discussed in~\cite{bbgr}. 

The structure of the mass terms can also be understood from Kaluza-Klein compactifications from 5 to 4 dimensions. 
Taking in $\bar\Psi_\mu \gamma^{\mu\rho\sigma} \partial_\rho \Psi_\sigma$ the index $\rho$ to be 5, 
and making the ansatz $\Psi_\sigma(x,x_5) \sim \sqrt{\gamma^5} \, \psi_\sigma(x) \, e^{i M x_5}$, one obtains the mass
term 
$M \bar\psi_\mu \gamma^{\mu\nu}  \psi_\nu$. 
In a similar manner, one may obtain the Fierz-Pauli mass term in~(\ref{eq_0}) from the Fierz-Pauli action~(\ref{eq_pf})
in 5 dimensions by setting $h_{\mu\nu}(x,x_5) = h_{\mu\nu}(x) e^{i M x_5}$ for $\mu,\nu = 0,3$. 

\vskip .5cm
{\bf Acknowledgments}\vskip .1in
\noindent
The work is supported  by NSF grants PHY-9722083, PHY-0070787, and PHY-9722101.

\end{document}